\documentclass[multphys,vecphys]{svmult}

\usepackage{makeidx}     
\usepackage{graphicx}    
\usepackage{multicol}    

\makeindex             

\begin{document}

\title*{Modeling Molecular-Line Emission from Circumstellar Disks}
\titlerunning{Modeling Line Emission from Disks}

\author{Michiel R. Hogerheijde}
\authorrunning{Hogerheijde}
\institute{Sterrewacht Leiden, P.O. Box 9513, 2300 RA, Leiden, The
   Netherlands
   \texttt{michiel@strw.leidenuniv.nl}}

\maketitle

\section{Introduction}
\label{s:intro}

Most young stars are surrounded by accretion disks
\cite{beckwith:nature}, which are likely progenitors of planetary
systems. The study of the physical structure, chemical composition,
and evolution of these disks is therefore clearly a topic of great
interest. The thermal emission of dust grains in disks has provided a
wealth of information on the disks' properties. At near-infrared
wavelengths, the thermal emission is dominated by a small amount
($\sim 10^{-5}$~M$_\odot$) of warm material close to the star, and the
emission is often optically thick. These wavelengths serve as
excellent probes of the presence or absence of a disk. At the longer
(sub) millimeter wavelengths, the emission probes the cold material at
larger radii ($>1$ AU), where the emission is usually mostly optically
thin. These wavelengths are good probes of the total mass of a disk.

Dust, however, only makes up 1\% of the total mass of the disk, if we
adopt the standard interstellar gas-to-dust mass ratio. Ignoring the
difficult to observe H$_2$ molecule (but see, e.g.,
\cite{thi:h2_co_disks, richter:h2disk}), circumstellar disks are now
commonly observed through lines of molecules such as CO, HCO$^+$, HCN,
CN, {\it etcetera\/}. Although it is found that the molecular
abundances are often strongly depleted through freezing out onto dust
grains (\S \ref{s:models}) and the lines correspondingly weaker,
molecular-line observations offer a powerful array of diagnostics to
study disks: from the relative intensities of transitions, the
molecular excitation can be derived and the density and temperature of
the gas constrained; from the line strengths, the molecular abundances
are found and the chemical composition inferred; and from the resolved
line profiles, the velocity field in the disk is directly probed. This
contribution discusses the methods with which this wealth of
information can be extracted from molecular lines, and presents
examples of recent research that illustrates this.

\section{The Physical and Chemical Structure of Disks}
\label{s:models}

Within the framework of standard accretion disks
\cite{lyndenbell:disks, pringle:araa}, recent theoretical models
specifically address their vertical structure \cite{dalessio:disk1,
chiang:disksed}. In a system where the central object dominates the
mass and the luminosity, the disk is expected to assume a flared
configuration, where the hydrostatic scale height increases with
radius \cite{shakura:bb_disk, kenyon:sedtts}. This flared surface
intercepts more stellar light, which raises the temperature of the
surface and, through reprocessing, of the disk interior (e.g.,
\cite{chiang:disksed}). Refinements to these models include the
effects of dust sublimation and self-shadowing of the outer disk by
inner, flared regions \cite{dullemond:shadow, dullemond:hole,
dullemond:vertical, dullemond:analysis2d}.

The resulting descriptions of the density and temperature as function
of radius and height then serve as input to calculate the abundances
of various molecules (e.g., \cite{aikawa:disk_co, aikawa:diskevol,
aikawa:disk_accretion, aikawa:disk2d, aikawa:deuterium_disk,
aikawa:2d_2, aikawa:warmlayers, willacy:photodisk,
markwick:innerdisks, bergin:ttsuv}; and Henning, this volume). Three
processes affect the molecular gas-phase abundances (reviewed in
\cite{evd:araa}): (1) freeze-out of molecules onto dust grains in the
dense and cold midplane, reducing the gas-phase abundances; (2)
evaporation of these ice mantles at intermediate vertical heights and
near the star, where the temperatures exceed the evaporation values
($\sim 22$~K for CO, $\sim 90$~K for H$_2$O) and where species return
to the gas; and (3) dissociation and ionisation by (inter-) stellar
ultraviolet photons in the disk surface where there is insufficient
shielding by dust, H$_2$, and CO, and where a chemistry akin to a
photon-dominated region is set up (see Sternberg, this volume). Radial
inflow and vertical convection further influence the
disk's chemical structure.

\section{Radiative Transfer and Molecular Excitation}
\label{s:radtrans}

With a model in hand for the physical and chemical structure of the
disk (density, temperature, systematic and turbulent velocity fields,
and gas-phase abundances -- all as functions of radius $R$ and height
$z$), the emergent emission (or absorption) lines can be
calculated. An iterative approach is usually chosen: Starting with an
educated guess for the level populations throughout the disk (or at a
sufficient number of grid points), the absorption and emission
coefficients are evaluated and the radiative transfer through the disk
is solved (along a sufficient number of photon paths). This yields the
intensity of the radiation field throughout the disk, with which the
equations of statistical equilibrium can be solved and an update to
the level populations found. This procedure is repeated until the
solution converges.

Several factors complicate this prescription in circumstellar
disks. Disks have a large range in density and temperature, with
molecular excitation ranging from thermodynamic equilibrium to
sub-thermal. Molecular abundances can vary by several orders of
magnitude within the disk. And opacities, both in continuum and in
lines, can be large. These factors require that the input model
contains sufficient detail to cover the variations in excitation and
abundance, and necessitate careful assessment of the convergence of
the solution: large opacities can strongly slow down the convergence,
because at each iteration step, changes in the population only
propagate over short distances (corresponding to $\tau \sim 1$).

Increased computer speed and memory has made realistic model
calculations tractable. In 1999 in a workshop at the Lorentz Centre in
Leiden, several of the leading codes for molecular excitation and
radiative transfer were (successfully!) tested against each
other\footnote{For more information, see {\tt
http://www.strw.leidenuniv.nl/$\sim$radtrans}.}. The reliability of
the obtained solutions depends on the quality of the molecular
collision rates that are used; several projects are under way to
provide accurate rates for astrophysically relevant species (see
Roueff, this volume; and Sch\"oier et al. in prep.\footnote{Available
through {\tt http://www.strw.leidenuniv.nl/$\sim$moldata} and {\tt
http://basecol.obs-besancon.fr}.}).

Slow convergence of iterative methods in the presence of large
opacities is not only an inconvenience for the impatient, but may lead
to incorrect results. Convergence is often judged on the difference
between subsequent iterations, with the implicit assumption that this
will be small only when the solution is close to the correct
value. The small differences between subsequent solutions due to
opacity is easily misinterpreted as convergence. In the modeling of
stellar atmospheres these problems are well understood, and have lead
to the development of acceleration techniques such as Accelerated
Lambda Iteration \cite{rybicki:ali_2} (ALI; Lambda Iteration refers to
iterative solution methods). Acceleration techniques have been
included in Monte Carlo approaches to solve the radiative transfer
\cite{mrh:code}\footnote{Available at {\tt
http://www.strw.leidenuniv.nl/$\sim$michiel/ratran.html}.}. This
combines the reliability of accelerated convergence with
flexibility in addressing arbitrary geometries.

\section{Examples}
\label{s:examples}

\emph{Resolved Disk Emission of LkCa~15.\/} Using the millimeter array
of the Owens Valley Radio Observatory (OVRO), the emission of the
$450$~AU radius, 0.18~M$_\odot$ disk around the young star LkCa~15 has
been resolved \cite{qi:phd, duvert:exdisk}. $^{12}$CO $J$=2--1
emission peaks at the source, with velocities consistent with
Keplerian rotation of an inclined ($i\approx 60^\circ$) disk around a
$0.8$~M$_\odot$ star. In stark contrast, the emission of CN 1--0 and
HCN 1--0 is broken up in two peaks, symmetrically displaced from the
star by $\sim 2'' \approx 300$~AU.

The appearance of HCN and CN can be reproduced by a simple radial
`step function' for the abundance, with CN and HCN only present
outside 300~AU \cite{kessler:aas2002}.  This can be
explained by chemistry in a flared disk, where only outside $\sim
300$~AU densities are low enough to allow ultraviolet radiation to
penetrate and set up a PDR with increased abundances of HCN and
CN. Self-consistent calculations are currently under
way (Kessler et al., in prep.).\\

\noindent\emph{Two-dimensional Transport of Ultraviolet Radiation.\/}
To accurately account for the penetration of ultraviolet radiation
from the central star into the flared disk, the full details of the
two-dimensional transfer must be included
\cite{zadelhoff:2duv}. Compared to one-dimensional calculations that
only include vertical transport, the predominantly forward-scattering
nature of the grains causes the ultraviolet field to penetrate
deeper. A larger fraction of the disk is subject to photo-chemistry,
with increased abundances of radicals like CN and C$_2$H.

An interesting application is the effect of the accretion-generated
ultraviolet radiation field on the HCN and CN
chemistry. Two-dimensional calculations show that the CN/HCN abundance
ratio increases by a factor of 10 if an ultraviolet excess similar to
that of TW~Hya is included on top of a 4000~K stellar blackbody
\cite{zadelhoff:2duv}. The ratio of the CN 3--2 line emission over HCN
4--3 increases by a similar amount, albeit confined to $\sim 1''$
region from the star at a typical distance of 140~pc for the
object. Instruments like the Smithsonian Submillimeter Array (SMA) and
the Atacama Large Millimeter Array (ALMA) will be able to trace such
regions.\\

\noindent\emph{Far-infrared $^{12}$CO Line Emission from Superheated
Disk Surfaces.\/} The object Elias~29 is a T~Tauri star with a face-on
disk of 0.012~M$_\odot$ that is obscured behind several layers of
interstellar clouds \cite{boogert:elias29}. It shows $^{13}$CO 6--5
and $^{12}$CO lines up to at least 20--19, indicating the presence of
$\sim 0.001$~M$_\odot$ of warm ($>200$~K) gas
\cite{ceccarelli:elias29}. This reservoir of warm gas may be explained
by material heated in outflow shocks, but is also comparable to the
amount of material expected in the superheated surface layer of the
flared disk around Elias~29.

Standard flared disk models \cite{chiang:disksed} cannot produce
sufficient far-infrared $^{12}$CO line emission: although the amount
of warm gas is sufficient, its density is too low to excite these
lines \cite{ceccarelli:elias29}. In the standard flared-disk model,
perfect mixing between gas and dust is assumed. If, however, the
grains have settled toward the midplane, the superheated layer extends
further into the denser disk interior. Calculations indicate that
$^{12}$CO far-infrared line strengths in this case can even exceed the
observed values, Although more detailed modeling is required,
including the fate of small dust grains that will not decouple from
the gas, this scenario suggests that the superheated layer is a viable
source of far-infrared CO emission.\\

\noindent\emph{Rotation and Infall in the Disk Around
L1489~IRS.\/} Submillimeter-continuum and millimeter aperture-synthesis
observations show that the object L1489~IRS is surrounded by a
flattened structure of $\sim 2000$~AU radius \cite{mrh:l1489} and
$\sim 0.02$~M$_\odot$ mass \cite{mrh:scuba}. While this mass is
entirely typical for disks around T~Tauri stars, its size exceeds by a
factor of 2--3 the largest known gas disks. Through modeling of the
HCO$^+$ 1--0 and 3--2 interferometer images, the velocity field in the
disk can be obtained \cite{mrh:l1489}. In addition to a Keplerian
component, suggesting a central mass of 0.65~M$_\odot$, inward motions
amounting to $\sim 10$\% of the total velocity vector are found. If
such motions continue inward to the star (the interferometer data only
sample scales $>700$ AU), the disk lifetime is $\sim 2\times
10^4$~yr. This is much shorter than the estimated duration of the
embedded (few times $10^5$ yr) and T~Tauri (few times $10^6$ yr)
phases. It suggests that L1489~IRS represents a short-lived
transitional stage, where the collapsing cloud core is settling onto a
rotationally supported disk.

The inward motions can be followed to within 0.1~AU from the star
through $^{12}$CO fundamental ro-vibrational absorption lines against
the stellar continuum at 4.7~$\mu$m \cite{boogert:l1489}. The line
profiles of the individual components of the P- and R- branches show
prominent wings extending to +100~km~s$^{-1}$. Such velocities are
expected to occur within 0.1~AU from the star, based on the velocity
model derived from the HCO$^+$ measurements above. The presence of
absorption line originating from rotational levels of $J$=6 and above
indicate temperature exceeding 100--200~K, which also are only
expected to be populated close to the star. Taken together, this shows
that the inward motions are present from 2000~AU as traced by the
HCO$^+$ data to within 0.1~AU probed by the CO absorption.

There are several objections against the interpretation that the
entire disk around L1489~IRS is contracting. The implied mass
accretion rate exceeds observational limits \cite{muzerolle:brgamma}
by more than an order of magnitude. And the theoretical models for
accretion flow predict subsonic motions, while the observed speeds
are supersonic. A possible solution is that only a thin, unstable
surface layer participates in the inflow, while the bulk of the disk
is Keplerian. Another solution posits the existence of two regions of
inflow: one at large radii (probed by the HCO$^+$ data) where the cloud's
material spirals onto the outer disk, and one at small radii (probed
by the CO line wings) where material streams from the disk onto the
star. The fact that both regions can be fit with the same velocity
model does not require them to be physically connected, because the
motions occur in the gravitational potential of the same star. Further
study of the L1489~IRS aims at distinguishing these scenarios
(Hogerheijde et al., in preparation).

\section{Conclusion}
\label{s:conclusion}

In this contribution I have shown that there now exist increasingly
realistic two-dimensional and three-dimensional models of the physical
and chemical structure of disks. Radiative transfer and molecular
excitation tools have been developed that can calculate the emergent
line spectrum on the basis of these disks models. On the observational
side, current data are beginning to probe the structure of disks,
although with no more than a few resolution elements across the
objects. The first results show interesting structural differences
between various probe molecules, which current chemical models can
explain.

In the next few years, increasing detail is to be expected in disk
models, as the increased speed of computers will bring more extensive
modeling within reach. With the arrival of new observational
facilities in the submillimeter (SMA, CARMA, ALMA) and the infrared
(SIRTF, SOFIA, Herschel), molecular-line modeling will form an
essential link in comparing models with data. By the end of this
decade, our understanding of protoplanetary disks will likely have
progressed significantly.\\

MRH wishes to thank the organisers of the conference and the Leids
Kerkhoven-Bosscha Fonds for their generous financial support.


%
%

%
%



\printindex
\end{document}